# Network medicine framework reveals generic herb-symptom effectiveness of Traditional Chinese Medicine

Xiao Gan[1,2,3,4,5†*], Zixin Shu[6†], Xinyan Wang[6], Dengying Yan[6], Jun Li[7], Shany ofaim[1,2], Réka Albert[4,5], Xiaodong Li[7,8], Baoyan Liu[9], Xuezhong Zhou[6*] and Albert-László Barabási[1,2,10*]

1. Network Science Institute, Northeastern University, Boston, MA 02115;

2. Department of Physics, Northeastern University, Boston, MA 02115;

3. Channing Division of Network Medicine, Department of Medicine, Brigham and Women's Hospital, Harvard Medical School, Boston, MA 02115

4. Department of Physics, Pennsylvania State University, University Park, 16802

5. Department of Biology, Pennsylvania State University, University Park, 16802

6. Institute of Medical Intelligence, School of Computer and Information Technology, Beijing Jiaotong University, Beijing, 100063, China

7. Hubei University of Chinese Medicine, Wuhan 430065, China.

8. Hubei Provincial Hospital of Traditional Chinese Medicine (Affiliated Hospital of Hubei University of Traditional Chinese Medicine, Hubei Academy of Traditional Chinese Medicine), Wuhan 430061, China.

9. China Academy of Chinese Medical Sciences, Beijing 100700, China.

10. Department of Network and Data Science, Central European University, Budapest 1051, Hungary

* To whom correspondence may be addressed: Dr. Xiao Gan, email: xxg114@psu.edu; Prof. Xuezhong Zhou, email: xzzhou@bjtu.edu.cn; Prof. Albert-László Barabási, email: a.barabasi@northeastern.edu.

† These authors contributed equally to the work

## Abstract

Traditional Chinese medicine (TCM) relies on natural medical products to treat symptoms and diseases. While clinical data have demonstrated the effectiveness of selected TCM-based treatments, the mechanistic root of how TCM herbs treat diseases remains largely unknown. More importantly, current approaches focus on single herbs or prescriptions, missing the high-level general principles of TCM. To uncover the mechanistic nature of TCM on a system level, in this work we establish a generic network medicine framework for TCM from the human protein interactome. Applying our framework reveals a network pattern between symptoms (diseases) and herbs in TCM. We first observe that genes associated with a symptom are not distributed randomly in the interactome, but cluster into localized modules; furthermore, a short network distance between two symptom modules is indicative of the symptoms' co-occurrence and similarity. Next, we show that the network proximity of a herb's targets to a symptom module is predictive of the herb's effectiveness in treating the symptom. We validate our framework with real-world hospital patient data by showing that (1) shorter network distance between symptoms of inpatients correlates with higher relative risk (co-occurrence), and (2) herb-symptom network proximity is indicative of patients' symptom recovery rate after herbal treatment. Finally, we identified novel herb-symptom pairs in which the herb's effectiveness in treating the symptom is predicted by network and confirmed in hospital data, but previously unknown to the TCM community. These predictions highlight our framework's potential in creating herb discovery or repurposing opportunities. In conclusion, network medicine offers a powerful novel platform to understand the mechanism of traditional medicine and to predict novel herbal treatment against diseases.

## I.    Introduction

Traditional Chinese medicine (TCM) prescribes combinations of natural medical products to treat disease according to the symptoms of a patient, offering a traditional yet personalized medicine approach for disease treatment[1-3]. While clinical data and studies of single herbs/prescriptions in the past years have shown the effectiveness of certain selected TCM-based treatments [4, 5], the general mechanistic nature of how TCM treat remains largely unknown. Two major challenges exist in investigating the mechanistic root of TCM: the first is the fact that TCM diagnosis, herb usage and prescription formation are all guided by classic TCM theory, which is not based on modern biology and medical science, resulting in a lack of scientific foundation to study TCM; the second challenge is complexity: each herb consists of dozens, if not hundreds, of chemicals with therapeutic potential; and each chemical can have many protein targets. Thus to understand TCM, it is imperative to establish a scientific framework/platform that can simultaneously connect TCM to modern bio-medical knowledge, and effectively work with the complexity of herb composition-target data.

Over the past two decades, novel methodologies and available data have offered new approaches to understanding traditional/natural medicine. A strategy to understand the therapeutic effect of a natural product is to leverage the multiple protein targets of its composing chemicals. Following this idea, a recent trend in TCM research has been the application of network pharmacology[6], which partially originated from the earlier drug-target network approaches developed by our lab[7-9]. Network pharmacology promotes the 'network target, multi-components' paradigm, complementing conventional research's limitation of focusing on single targets. This approach has helped TCM researchers identify herbal chemicals with therapeutic potentials, better understand mechanisms of action, and promote new drug discovery [10-12]. However, existing TCM network pharmacology approaches are limited to single herbs or single prescriptions, missing a high-level perspective of the totality of TCM herb-disease relations. Moreover, the network pharmacology approaches only consider herbs/drugs that target disease genes directly, not considering any network propagation effect, e.g. when perturbing a target has further downstream/cascading effect through protein interactions. Therefore, the approaches used so far do not use the underlying network connections between herb/drug targets and disease genes to their full extent. Here we propose avenues to overcome these limitations.

Recently, the network medicine framework has successfully revealed the general patterns of diseases, drugs and their relations on the human protein-protein interactome (PPI), a network whose nodes are proteins that link to each other by physical (binding) interactions [13]. Disease-associated proteins tend to form locally clustered modules in this network, and shorter network distance between two disease modules is indicative of their comorbidity [14]. Furthermore, one can predict drug efficacy by leveraging the network relation of drug targets with disease modules [15, 16], leading to powerful drug-repurposing methodologies [17]. In our most recent work, we predicted effective drugs for treating COVID using network-based methodologies and discovered the network patterns of the effective drugs [18]. Network medicine as a platform not only enables high-level, systematic study of herb-disease relations, but also allows the characterization of network cascading effects from disease genes and drug targets, using the underlying protein interactions in the protein interactome network.

In this work we ask a simple but fundamental question: is there a general principle that can explain overall how TCM herbs works? To answer this question, we develop a generic network medicine framework to characterize TCM as the relation between herbs and disease symptoms, which manifest in the human protein interactome as the network-based relation between herb targets and symptom-associated genes. We build our framework on symptoms instead of diseases because TCM diagnosis and herbal therapies are all based on patient symptom phenotypes rather than disease diagnosis. By studying relations between symptoms and herbs on the interactome, we discover that: (I) genes associated with a symptom tend to cluster into a local interactome module; moreover, shorter network distances between symptom modules are indicative of symptom similarity and co-occurrence. (II) The network proximity between an herb's targets and a symptom module is indicative of the herb's effectiveness in treating the symptom. We validate our network medicine framework with empirical data and real-world clinical data, and highlight its application in identifying novel herb discovery/repurposing opportunities. We present our study design in Figure 1.

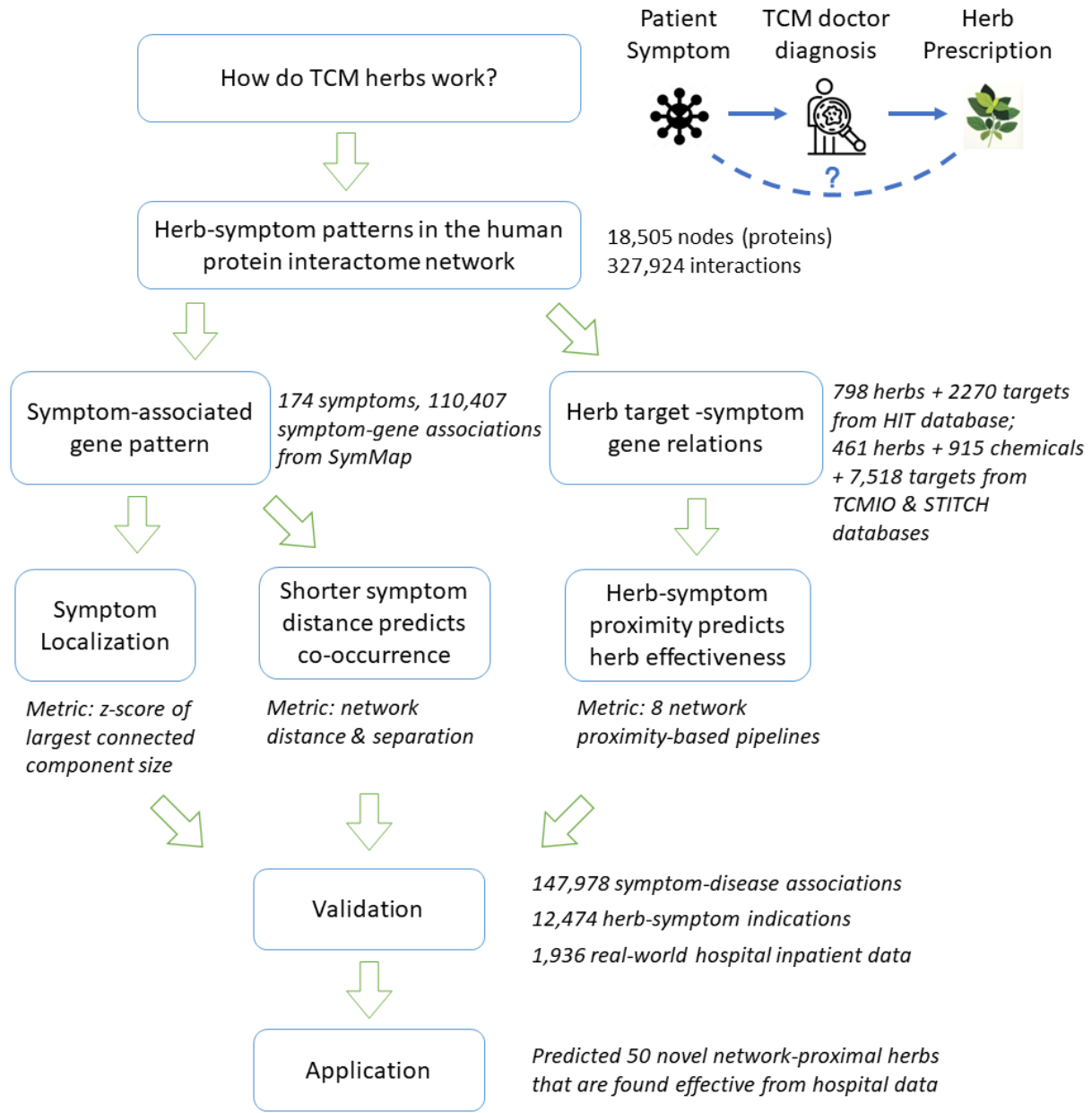

*Figure 1. Study design of our work. We show that Traditional Chinese medicine can be characterized as symptoms and herb relation, in the human protein interactome. We discover the network pattern connecting symptoms and herbs, then validated these relations with empirical data and real-world hospital data. We highlight the application of our work in predicting novel herb-symptom treatments*

## II. Symptom-associated genes form modules in the protein interactome, and the network distance of two symptom modules is indicative of their similarity

One of the main challenges to understand TCM is how to connect the classic TCM concepts to modern bio-medical knowledge, e.g. TCM does not have the "disease" concept of modern medicine. To overcome this issue, we leverage the fact that TCM clinical diagnosis is based on patient symptom phenotypes, and that TCM herbal therapies are designed to treat symptom phenotypes too. In other words, TCM diagnosis and treatment is similar to modern medicine and clinical diagnosis, if one considers symptom phenotypes instead of diseases [19, 20]. We therefore propose to characterize the indication and effect of TCM by symptoms, which connects classic TCM practice with modern bio-medical knowledge. Moreover, symptoms are well-supported by disease taxonomy[21, 22] and gene association data, providing abundant data foundations for our study. In summary, we propose to characterize TCM with symptoms; we study the network pattern of symptoms' associated genes in the interactome in this section, and will study symptom-herb relation in the next section.

We characterize each symptom by its associated genes, which capture the underlying biological mechanisms of the symptom. These genes are obtained from a curated symptom-gene association dataset from a previous work (see Methods) [23]. Then we project these genes onto the human interactome, and focus on 174 symptoms with at least 20 associated genes, assuming the symptoms with fewer genes have too incomplete gene data, similar to our previous approach [14]. We find that for 108 of these 174 symptoms, the genes associated with each symptom form a significantly larger connected component than random expectation (z>1.6, Figure 2A), implying that the genes associated with a symptom tend to agglomerate into a localized module in the interactome, rather than being distributed randomly. Furthermore, genes associated with different symptoms are distant from each other (Figure 2B), characterized by the network separation metric (see Methods), with avg. $S_{ab}$= 0.23, larger than the random expectation of zero. This suggests different symptoms perturb different regions of the protein interactome.

We further ask if the network distance between different symptoms modules on the interactome can reveal relations between the symptoms. First, we investigate symptoms' co-occurrence in diseases: we leverage the 147,978 symptom-disease association data from our previous work [24], and computed how many shared diseases two symptoms are associated with, as the co-disease count of two symptoms. We found that the average network distances between two symptoms' associated gene modules ($D_{ab}$) negatively correlate with their co-occurrence in diseases (Figure 2D). Next we investigate the biological similarity of symptoms: we leverage the GO semantic similarity (see Methods) [25] of genes, which characterizes the similarity of two genes based on their similarity in GO annotations. For two symptom modules, we compute the average GO semantic similarity between their associated genes as the symptoms' biological similarity, and find it to negatively correlate with network distance $D_{ab}$ as well (Figure 2E). Together, these findings indicate that the closer the two symptom modules are in the protein interactome, the more likely they will co-occur in the same disease, and the more biologically similar they are. For example, the symptom pair fever and diarrhea has a short average network distance $D_{ab}$=1.25, and a high co-disease count of 1278 (Figure 2D). Other frequently co-occurring symptom pairs include fatigue-pain ($D_{ab}$=1.25, co-disease count = 1163) and dizziness-headache ($D_{ab}$ = 1.32, co-disease count = 917). For diarrhea and fever, they co-occur in many diseases such as inflammatory diseases (e.g. inflammatory gastroenteropathy)[26] and virus induced infectious diseases (e.g. severe acute respiratory syndrome coronavirus 2 [27]). These co-occurrences may be rooted in the two symptoms' 27 shared genes, including inflammatory biomarkers (e.g. PIK3R1 and TNF[28]) and the cytokines (e.g. IL1A and IL7R[29]). Their associated pathways tend to be related to the inflammatory immune processes, such as the JAK/STAT pathway [30] and cytokine-mediated signaling pathway[31]. On the other hand, symptoms with higher network distance tend to be different with less co-occurrence, for example eye pain and anorexia ($D_{ab}$ = 2.91, co-disease count: 13).

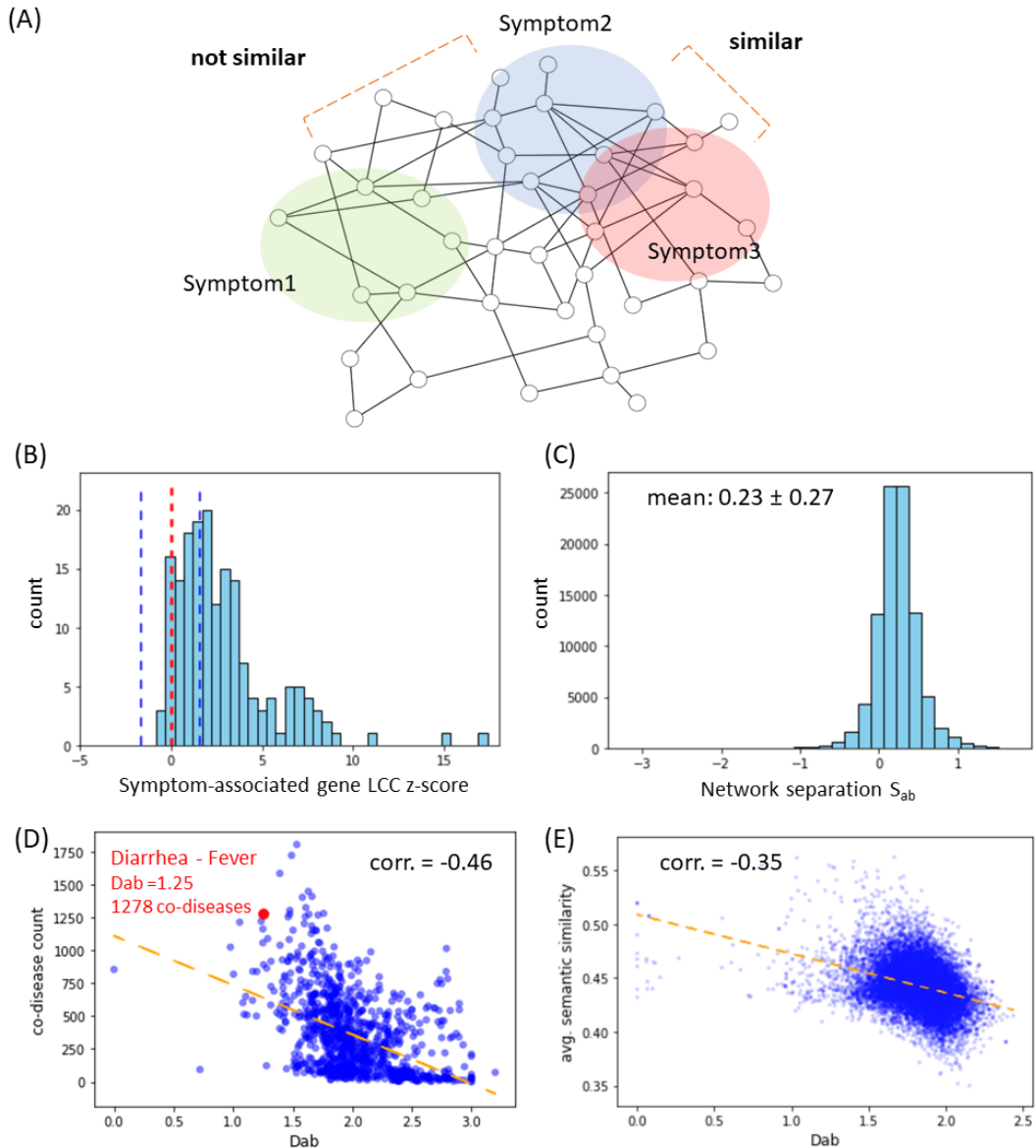

Figure 2. Symptom pattern in the human protein-protein interactome (PPI),. (A) Schematic showing that genes associated with a symptom form localized modules on the network, and the inter-module network

*distance is indicative of symptom similarity. (B) Distribution of largest-connected-component z-score formed by symptom-associated genes, for 174 symptoms. 108 out of 174 symptoms form significantly clustered local modules (z>1.6). The blue dotted lines indicate z=±1.6, and the red dotted line for z=0. (C) Distribution of network separation ($S_{ab}$) of all symptom pairs. The average <$S_{ab}$> is larger than zero, the random expectation. This suggests that different symptoms perturb different/specific regions in the PPI, by forming modules distant from each other. (D) the average interactome network distance ($D_{ab}$) of a symptom pair negatively correlates with the symptoms' co-occurrence in diseases (co-disease count). Each dot represents a symptom pair. An example of diarrhea and fever is highlighted in red. (E) the interactome network distance of a symptom pair negatively correlates with the biological similarity of the genes associated with the symptoms.*

### III. Herb-symptom network proximity is indicative of the herb's therapeutic effect against the symptom

We characterize TCM herbs by their protein targets in the human protein interactome. However, the challenge of complexity is that an herb is composed of many chemicals, and each chemical can bind to multiple protein targets; there is no prior knowledge on how to optimally define an herb's targets from its composing chemicals. To overcome this challenge, we design a multimodal approach to characterize the network relation between an herb and a symptom module, by integrating multiple datasets and developing multiple representative network metrics. We use two groups of datasets: (1) We obtain pre-defined herb target data from the recently updated HIT database; (2) We obtain herb chemical composition data TCM databases and integrate with chemical target data from STITCH. For (1), we directly use herb target data from the recent HIT 2.0 database[32], where the curators text-mined literature abstracts for compound-target relations, followed by manual review. After mapping the herbs and targets to our herb name data and protein interactome, the HIT 2.0 database yields 798 herbs and 2270 targets, with an average of $162.9 \pm 185.5$ targets per herb. For (2), we use the herb chemical composition data from TCMIO database, which is a recent and comprehensive collection of the TCMSP, TCMID, and TCM-ID databases [33-35]. These databases focus on chemicals with a potential therapeutic effect, rather than including the full chemical composition of a herb. We then use STITCH chemical target data, keeping only targets with experimental evidence to ensure reliability. In the end, we arrive at a total of 461 herbs with target data, 915 chemicals that have target data and appear at least in one herb, and 7,518 unique protein targets. On average, each herb has $61.9 \pm 61.5$ chemicals, and each chemical has $69.7 \pm 311.4$ targets.

Next, to characterize herb-symptom network relation, we develop a multimodal approach that yields a network-based metric for each herb-symptom pair (see schematics and workflow in Figure 3A, B). Our hypothesis is that herbs effective for treating a symptom are targeting proteins proximal to symptom-associated genes in the interactome, similar to the network pattern of drug-disease relations[15]. We use two proximity measures, (i) proximity distance and (ii) proximity z-score, to characterize the network relation between a set of targets and a set of symptom-related genes (see Methods). The proximity distance is the average distance between herb targets to their closest symptom-associated gene(s); and the proximity z-score measures how the proximity distance differs from random expectation, with z=0 being neutral, z<0 being proximal than random, and z>0 being more distant from random. For both metrics, the lower the metric value, the closer the network relation is. HIT data directly associate targets to each herb, so we compute the two proximity measures straightforwardly as a herb-symptom metric (herb-target mapping method (a)). On the other hand, the TCM herb-chemical-target dataset does not have direct herb-target associations, so no direct herb-symptom relation metric can be applied. Therefore, we design three additional herb-target mapping methods (b-d) to obtain herb-symptom network metrics: (b) Target Union: we define an herb's target set as the union of the targets of all composing chemicals of the herb, then we interpret herb target – symptom proximity measures as herb-symptom relation metrics. (c) & (d): we define a $2^{nd}$ order herb-symptom distance from $1^{st}$ order chemical-symptom distances: first, for every chemical-symptom pair, we compute chemical-symptom proximity metrics using the chemical's targets; then we define the $2^{nd}$ order herb-symptom distance as: (c) the average of all chemical-symptom distances from the herb's composing chemicals; or (d) the smallest of all chemical-symptom distances from the herb's composing chemicals (see Methods). The mechanistic assumption of (a) and (b) is straightforward; the mechanistic assumption of (c)&(d), i.e., of the $2^{nd}$ order distances, is to approximate the combinatorial effect of multiple chemicals composing a herb: (c) takes the average, assuming the effect of a herb is the average of its composing herbs; (d) takes the minimum, assuming the most proximal chemical dominates the herb's effect. Altogether, the two distance/proximity metrics (proximity d & z) and the four herb-target mapping methods (a-d) give us 8 pipelines to compute herb-symptom network metrics (Figure 3B), for each herb-symptom pair. We assemble a dataset with the herbs with network metrics from all 8 pipelines and provide it in the supporting data.

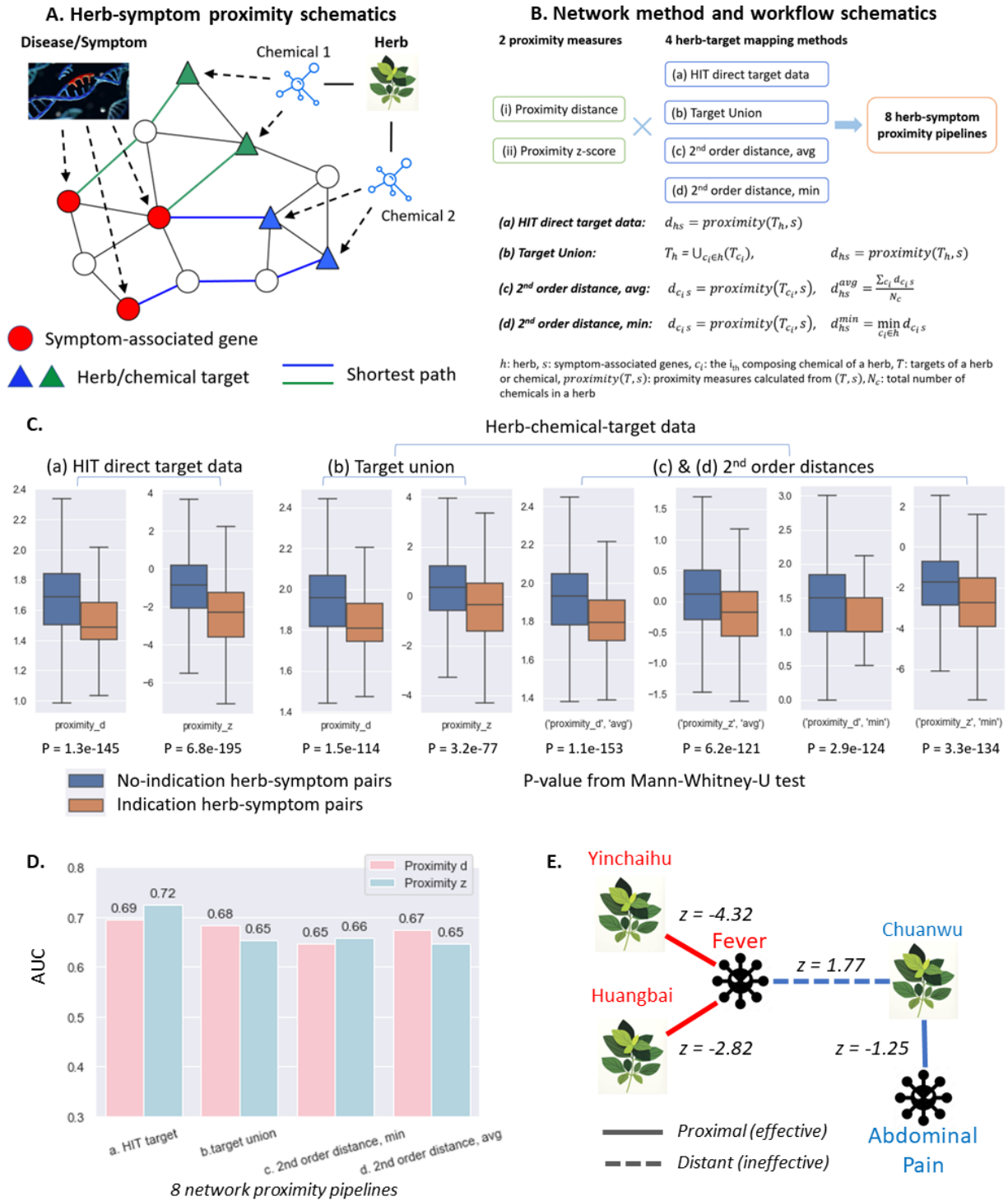

*Figure 3. Herb targets and herb-symptom proximity. (A) Schematics of the herb-symptom network proximity metric, based on shortest paths between herb-chemical targets and symptom associated genes in the protein interactome; (B) Workflow of the multimodal approach for 8 herb-symptom proximity pipelines, with quick definitions of the metrics; (C) Results of the 8 pipelines of network metrics for herb-symptom pairs categorized as indicated or non-indicated. Indicated herb-symptom pairs (orange bars) show lower proximity metrics (shorter network distance) consistently over all 8 pipelines; (D) AUROC*

*performance evaluation of the 8 herb-symptom proximity pipelines, using the known herb-symptom indications as positive cases. (E) Example demonstrating herb-symptom proximity: herbs Yinchaihu and Huangbai are proximal to the Fever symptom and are indeed used to treat fever in practice; whereas the Chuanwu herb is distant to Fever but proximal to Abdominal pain, thus it is not used to treat Fever but to treat Abdominal pain.*

To evaluate the performance of our network metrics, we leverage a dataset of expert-curated herb-symptom indications from SymMap[36] as ground truth, where the herb is recognized by Chinese Pharmacopoeia (CHPH, the most authoritative TCM data, 2015 edition) to be effective against the symptom. We map these herb-symptom indication pairs into our 8 pipelines, resulting in 1,480 indications in HIT pipelines (a), and 1,325 indications in the Target Union and 2nd order distance pipelines (b), (c) & (d). To evaluate our network proximity hypothesis, first we make box-and-whiskers plots of the network metrics for all 8 pipelines (Figure 3C), comparing indicated herb-symptom pairs (orange bars) against non-indicated herb-symptom pairs (blue bars). We observe the orange bars are lower than the blue bars consistently over all pipelines, showing that known effective herb-symptom pairs are more proximal, compared against other herb-symptom pairs. Alternatively, we calculate the AUROC scores for all pipelines, using the same herb-symptom pairs with indication as positive cases, and the herb-symptom pairs without indication as negative cases (Figure 3D). Again, we find that all pipelines are predictive (AUC>0.5). Both proximity distance and the proximity statistical z-score are predictive, and neither performs consistently better than the other one. The best-performing pipeline is the HIT target dataset with proximity z-score, reaching the highest AUC, 0.72. To our knowledge there's no similar generic method that predicts a herb's effectiveness against a symptom/disease form the protein interactome on a large scale, so we compare this result with our previous results on drug-disease relations. The observed best 0.72 AUC value here is higher than the best performances in previous works, namely the best AUC=0.66 in generic drug-disease effectiveness [15], and the best AUC = 0.63 in drug-COVID effectiveness [18]. We conclude that the (relatively) high AUCs and the consistency over all pipelines shows that network proximity has predictive power regarding TCM herb-symptom effectiveness. This result is especially remarkable considering the high noise in such large-scale and multi-faceted data, and the diversity of disease-herb relationships. The good performance of HIT target data may have risen from their very recent update, with well-curated target data.

We further demonstrate herb-symptom proximity with known examples, using the best performance HIT z-score pipeline (Figure 3E). Taking the "Fever" symptom for example, herbs with highly negative network proximity z-scores to fever include *Yinchaihu* (*Radix Stellariae*, Starwort root, z-score: -4.32), which treats asthenic fever in the late stage of febrile diseases[37] by regulating a series of inflammatory processes, such as nuclear factor (NF)-dB and mitogen-activated protein kinase (MAPK)[38]. Another herb that treats fever, *Huangbai* (*Phellodendri Chinensis Cortex*, Phellodendron Bark, z-score: -2.82), is also used to treat various diseases characterized by fever, such as pneumonia and tuberculosis[39]. Berberine, one of the main active chemical components of *Huangbai*, has an anti-inflammatory and antipyretic effect[40]. On the other hand, a herb distant from the fever symptom in the interactome, such as *Chuanwu* (*Radix Aconitum*, Aconite Root, z-score: 1.77), is unlikely to be effective against fever, consistent with expert knowledge. *Chuanwu* is network-proximal to abdominal pain (z-score: -1.25), and indeed is used for pain relief for its anti-inflammatory, analgesic and anti-tumor effects[41].

### IV.    Validation of network-based herb-symptom relation with hospital data

In this section we validate the symptom relations and the herb-symptom proximity of our network medicine framework with real-world hospital data. We collected the electronic medical record data of 1936 Liver cirrhosis inpatient cases from Hubei Provincial Hospital of Traditional Chinese Medicine in Wuhan. Information on symptoms and their changes (before and after treatment) were extracted from the admission and discharge records using a clinical information extraction tool (Human-machine Cooperative Phenotypic Spectrum Annotation System, www.tcmai.org, HCPSAS)[42]. Since the symptoms in the clinical data are in Chinese, we manually mapped them to UMLS terms to enable symptom-gene association data (see Methods). Similarly, we map herbs from their Chinese names in the data to herbs IDs for their chemical and target data. In the end, the hospital data contain a total of 114 symptoms, 218 herbs, and 23,413 herb-symptom pairs.

First, we validate the relation between a symptom pair's network distance and their co-occurrence. We leverage the symptom appearance frequency in the inpatient data, and compute the relative risk (RR) between each symptom pair as their co-occurrence metric. Relative risk is a standard statistic that measures the strength of an association (in this case, co-occurrence of two symptoms), defined as ratio of the probabilities of the exposed and unexposed groups. Then we compute the network distance between each pair of symptoms (See supporting info). Finally, for all symptom pairs, we compare their relative risk with their network distance, and find a Pearson correlation of -0.31 (Figure 4A). This negative correlation validates our hypothesis that if two symptoms have shorter network distance, they are more likely to co-occur. Specifically, examples of symptom pairs include nausea and vomiting ($D_{ab}$ = 0.53, RR = 11.00) as well as consciousness disorder and lethargy ($D_{ab}$ = 0.62, RR = 10.6). Conversely, symptom pairs with longer network distance do not have high relative risk, for example disorder joint and poor appetite ($D_{ab}$ = 1.75, RR = 0.98) and abdominal distension-ulcer mouth (Dab: 2.00, RR:0.87).

Second, we show that network proximity captures doctors' knowledge in prescribing herbs against symptoms, by comparing the network proximity of herb-symptom pairs in the clinical dataset (i.e. herbs prescribed by doctors) against that of herb-symptom pairs absent from the clinical dataset (i.e. herbs not prescribed by doctors). This will tell us whether the prescribed herbs against symptoms are different (in a network sense) from herbs that are not prescribed. We observe in Figure 4B that, for all 8 proximity pipelines, the herb-symptom pairs in the hospital dataset (orange boxes) have significantly lower network proximity metric (i.e. they are closer) than the herb-symptom pairs not in the clinical dataset (blue boxes). In other words, doctors prescribe herbs proximal to the disease/symptoms in their practice. This supports our hypothesis that proximal herbs are more likely to be effective, by showing that network proximity is consistent with doctors' expert knowledge.

Next, we show that network proximity predicts effective herb-symptom pairs found in the clinical dataset. As the data do not contain any metric of herb effectiveness, we need to start by defining a herb effectiveness metric. To ensure data reliability, we focus on a subset of the data with representative symptoms and sufficient data to support bioinformatics analysis with statistical significance. We apply propensity score matching (PSM) method to 888 herb-symptom pairs with a frequency of at least 30, i.e. there are at least 30 cases where the patient with this symptom is treated with the herb. Propensity score matching means that for each herb-symptom pair, we matched the patients with this symptom and treated with this herb (i.e. the case group), to a control group where the patients have the same symptom but are not treated with this herb; and we adjusted for potential confounders (e.g. age, gender, history of hypertension, diabetes, coronary artery disease, chronic kidney disease) of the patients in the control group (See Methods, and an example in the next paragraph). After PSM we identified 86 herb-symptom pairs where the case group has significantly higher symptom recovery rate than the control group ($p < 0.05$, chi-square test), i.e. the herb treatment is effective by statistical significance. We show that these 86 effective herb-symptom pairs are network-proximal, compared against all herb-symptom pairs (Figure 4C). Note that since we showed the herb-symptom pairs used in the clinical dataset are already more proximal than random (Figure 4B), the clinical dataset is a biased dataset (over-filled with positive/effective herbs). For this reason, we cannot compare the effective pairs against other pairs within the clinical dataset, as they are both positive samples. Instead, we have to compare the effective pairs against all herb-symptom pairs with network proximity metrics. As shown in Figure 4D, all 8 pipelines showed that the effective pairs have significantly lower network metrics, validating that network proximity is indeed a good predictor of herb effectiveness in real-world hospital data.

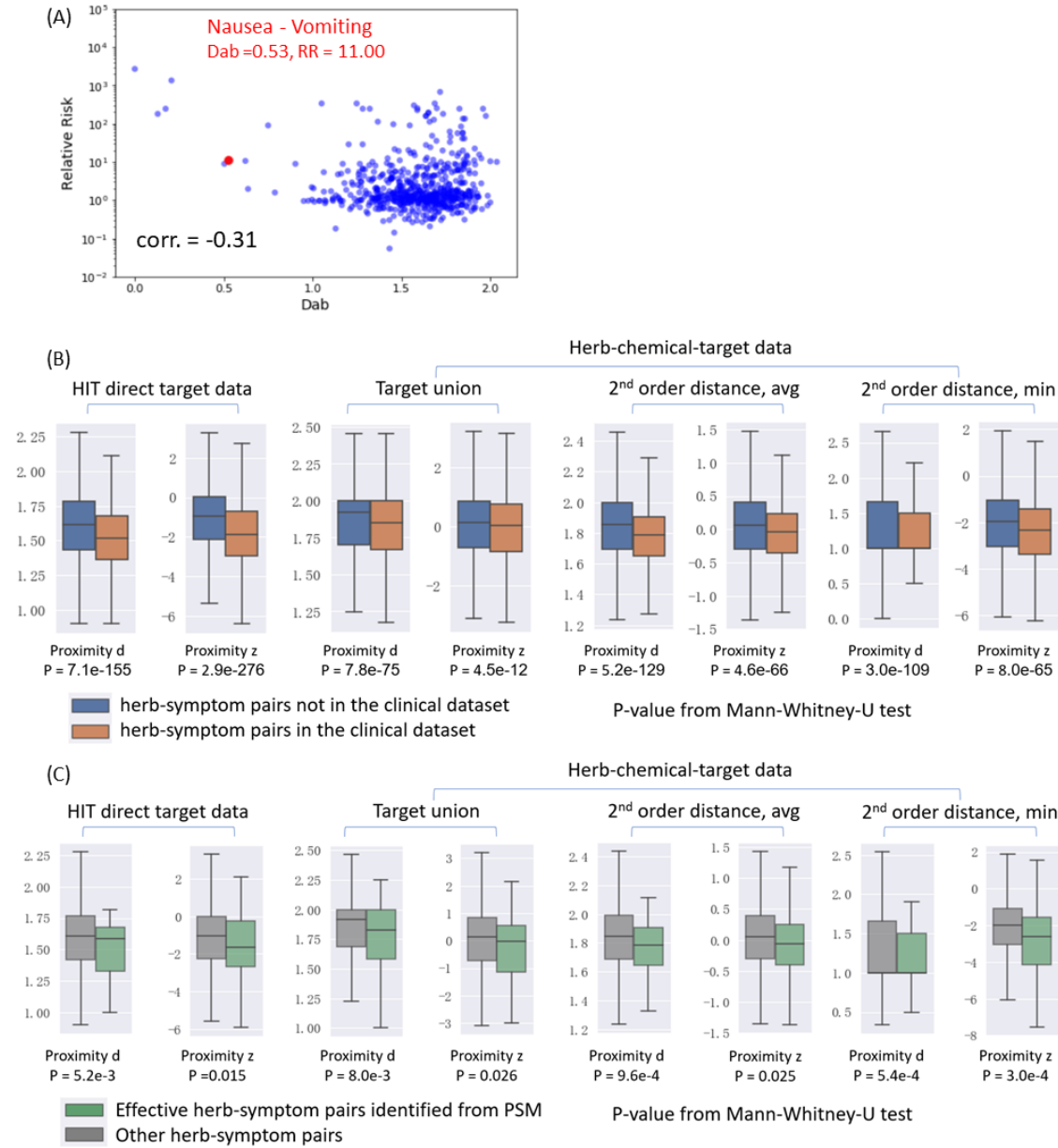

*Figure 4. Validation of network medicine framework with hospital data. (A) Patient symptom data shows a negative correlation between symptom pair relative risk (in log scale) and network distance Dab, validating the network-based prediction in real hospital patient data, indicating shorter network distance is predicative of higher symptom co-occurrence; (B) Herbs used by doctors in clinical data (orange boxes) are significantly more proximal to symptoms than herbs not used in clinical data (blue boxes), consistently observed over all 8 pipelines; (C) The 86 effective herb-symptom pairs identified from propensity score matching (orange boxes) have lower network metrics than other herb-symptom pairs (blue boxes), i.e. network proximity metrics can predict the significantly effective herb-symptom pairs, consistently in all 8 pipelines.*

We demonstrate our methodology with the example of the herb-symptom pair *Baizhu*[1] - poor appetite. Network proximity[2] shows a negative z-score=-2.45 between *Baizhu* and poor appetite, meaning the herb's target hits proximal to the symptom's associated genes, suggesting *Baizhu*'s potential effectiveness in improving poor appetite. To evaluate from the patient data the effectiveness of the herb on the symptom, we match patients with poor appetite that are treated with *Baizhu* (case group), to patients with poor appetite but not treated with *Baizhu* (control group), and compare their symptom recovery rate. We observe that in the matched patients, *Baizhu* significantly improved the recovery rate of poor appetite (79.53% case group recovery rate v.s. 72.51% control group recovery rate, P-value = 0.0316), consistent with the network proximity prediction. Indeed, *Baizhu* is a medicinal plant used to treat gastrointestinal dysfunction according to the Chinese pharmacopoeia. Studies showed that atractylenolide I, sourced from Baizhu, regulates gastrointestinal function and promotes the absorption of nutrients[43], supporting the effectiveness of Baizhu in improving a patient's appetite.

### V.  Our network medicine framework reveals herb discovery and repurposing opportunities

A highlight of our network medicine framework is predicting novel herb candidates to treat symptoms, using network proximity metrics as a predictor of a herb's effectiveness against a symptom. Thus, the herb-symptom pairs with highly negative network proximity metrics are promising predicted candidates. For example, "*Chaihu*[3]-abdomen distention" is a herb-symptom pair not recognized by Chinese pharmacopoeia as effective; however, we found a negative proximity z-score=-2.86 between the herb and the symptom, predicting the herb as potentially effective against the symptom. We found validation of this prediction in the clinical dataset that (1) *Chaihu* is frequently prescribed in practice to treat abdomen distention (their herb-symptom co-occurrence frequency is 381, significantly higher than the average frequency of 106.8 ± 106.5 for all herb-symptom pairs); and (2) in the PSM matched patients, *Chaihu* significantly improved the recovery rate from abdomen distention (88.71% vs 83.73%, P-values =0.0458). Indeed, *Chaihu* contains chemicals such as saikosaponins, which can relieve abdomen distention caused by dyspepsia and ascites of liver cirrhosis[44], suggesting its potential effectiveness against abdomen distention. Furthermore, we found effective herb-symptom pairs less frequent in clinical practice, such as "*Cangzhu*[4]- abdominal pain" pair, which is predicted effective by a negative proximity z-score = -3.08 and has a significantly improved recovery rate (93.55% vs 70.97%, P-values =0.0461,). Studies have shown that the volatile oil, component of *Cangzhu*, has anti-acetylcholine effect, which can relieve abdominal pain symptoms caused by intestinal spasm[45]. More interestingly, we also found potentially effective herb-symptom pairs that are rarely reported in any studies, such as the "*Baiji[5]*-edema" pair, which has a highly negative proximity z-score of -4.12 with improved recovery rate (83.33 % vs 67.86%, P-values=0.0195). This suggests that *Baiji*, an astringent hemostatic conventionally used to relieve gastrointestinal bleeding[46], might be effective in relieving edema.

---

[1] *Baizhu*: *Atractylodis Macrocephalae Rhizoma,* Rhizome of Largehead Atractylodes
[2] We use HIT proximity z-score throughout this section and the next section in our example demonstrations, because it was the best-performing pipeline found from previous section.
[3] *Chaihu*: *Bupleuri Radix*, Root of Chinese Thorowax
[4] *Cangzhu: Atractylodis Rhizoma,* Rhizome of Swordlike Atractylodes

The findings of potentially effective herb-symptom treatments highlight the predictive power of our framework in identifying herb discovery or repurposing candidates. We provide in Table 1 a list of 50 herb-symptom pairs that are both network proximal and effective after PSM in clinical data, but not yet recorded in the Chinese Pharmacopoeia. These are promising new treatment candidates where the herb's effectiveness against the symptom or disease is not yet recognized but can be tested in follow-up studies.

*Table 1. 50 herb-symptom pairs with negative network proximity z-score (i.e. predicted as potentially effective), and found effective in propensity score matched clinical data. They are promising candidates for novel herb-symptom treatment discovery/repurposing. The table is ordered from the most negative proximity z-score to the least negative. The $3^{rd}$ column is the network proximity z-score; the $4^{th}$ column is the number of patients in the case/control group after propensity score matching; the $5^{th}$ and $6^{th}$ are the recovery rates of the case group and the control group; the last column is the p-value for the recovery rate difference, from a chi-square test.*

| Herb | Symptom | Proximity Z-score | Number of patients | Case Group Recovery Rate | Control Group Recovery Rate | P-value |
|------|---------|-------------------|--------------------|--------------------------|-----------------------------|---------|
| *Bei Sha Shen* | Edema | -6.31 | 79 | 77.22% | 58.23% | 1.07E-02 |
| *Jin Yin Hua* | Edema | -5.58 | 78 | 73.08% | 53.85% | 1.26E-02 |
| *Hu Ji Sheng* | Edema | -5.31 | 46 | 91.30% | 69.57% | 1.80E-02 |
| *Xiang Fu* | Edema | -4.32 | 63 | 73.02% | 53.97% | 2.64E-02 |
| *Chi Shao* | Edema | -4.30 | 98 | 80.61% | 68.37% | 4.93E-02 |
| *Bai Ji* | Edema | -4.12 | 84 | 83.33% | 67.86% | 1.95E-02 |
| *Ku Shen* | Abdomen distention | -3.37 | 40 | 92.50% | 72.50% | 3.94E-02 |
| *Chen Pi* | Body pain | -3.16 | 87 | 93.10% | 81.61% | 2.25E-02 |
| *Cang Zhu* | Abdominal pain | -3.08 | 31 | 93.55% | 70.97% | 4.61E-02 |
| *Xiang Fu* | Fatigue | -2.96 | 201 | 87.56% | 80.10% | 4.21E-02 |
| *Mu Xiang* | Fatigue | -2.90 | 149 | 92.62% | 85.23% | 4.23E-02 |
| *Chai Hu* | Abdomen distention | -2.86 | 381 | 88.71% | 83.73% | 4.58E-02 |
| *Sha Ren* | Poor appetite | -2.77 | 112 | 81.25% | 67.86% | 2.14E-02 |
| *Zhe Bei* | Abdomen distention | -2.76 | 64 | 95.31% | 81.25% | 2.79E-02 |
| *Chen Pi* | Fatigue | -2.67 | 579 | 87.39% | 83.25% | 4.63E-02 |
| *Shan Zha* | Abdominal pain | -2.67 | 53 | 81.13% | 62.26% | 3.11E-02 |
| *Zhi Zi* | Poor appetite | -2.48 | 73 | 84.93% | 71.23% | 4.54E-02 |

---

[5] *Baiji: Bletillae Rhizoma*, Bletilla Striata Rchb.F.

| Fang Ji | Abdomen distention | -2.36 | 85 | 95.29% | 84.71% | 4.08E-02 |
|---|---|---|---|---|---|---|
| Yi Zhi Ren | Fatigue | -2.34 | 38 | 92.11% | 68.42% | 2.11E-02 |
| Shan Yao | Edema | -2.29 | 150 | 80.67% | 70.00% | 3.21E-02 |
| Fang Feng | Abdomen distention | -2.27 | 67 | 91.04% | 76.12% | 1.97E-02 |
| Zhe Bei | Poor appetite | -2.21 | 67 | 86.57% | 71.64% | 3.36E-02 |
| Fang Feng | Abdominal pain | -2.16 | 30 | 83.33% | 60.00% | 4.49E-02 |
| Chai Hu | Fatigue | -2.15 | 593 | 88.20% | 82.63% | 6.63E-03 |
| Lian Qiao | Cough | -1.94 | 56 | 89.29% | 75.00% | 4.84E-02 |
| Xuan Shen | Poor appetite | -1.80 | 64 | 82.81% | 67.19% | 4.12E-02 |
| Zhi Shi | Fatigue | -1.77 | 399 | 88.47% | 79.95% | 9.64E-04 |
| Hong Hua | Poor appetite | -1.73 | 37 | 89.19% | 67.57% | 4.81E-02 |
| Tai Zi Shen | Insomnia | -1.72 | 234 | 87.18% | 77.35% | 5.38E-03 |
| Ban Bian Lian | Fatigue | -1.68 | 86 | 86.05% | 72.09% | 2.45E-02 |
| Yu Jin | Abdomen distention | -1.59 | 351 | 90.31% | 84.90% | 2.95E-02 |
| Fang Feng | Insomnia | -1.56 | 55 | 90.91% | 72.73% | 1.34E-02 |
| Hui Xiang | Fatigue | -1.49 | 31 | 93.55% | 70.97% | 4.61E-02 |
| Hu Ji Sheng | Poor appetite | -1.36 | 45 | 86.67% | 64.44% | 1.42E-02 |
| Chuan Qiong | Poor appetite | -1.35 | 146 | 80.82% | 68.49% | 1.54E-02 |
| Zhi Mu | Poor appetite | -1.32 | 102 | 82.35% | 70.59% | 4.76E-02 |
| Hu Ji Sheng | Insomnia | -0.84 | 34 | 94.12% | 67.65% | 1.36E-02 |
| Hou Pu | Fatigue | -0.71 | 369 | 86.45% | 80.22% | 2.31E-02 |
| Zhi Mu | Insomnia | -0.69 | 85 | 91.76% | 77.65% | 1.06E-02 |
| Xuan Fu Hua | Fatigue | -0.68 | 93 | 89.25% | 77.42% | 3.04E-02 |
| Xu Zhang Qing | Fatigue | -0.54 | 35 | 94.29% | 71.43% | 2.64E-02 |
| Suan Zao Ren | Poor appetite | -0.44 | 187 | 84.49% | 75.94% | 3.78E-02 |
| Huang Lian | Fatigue | -0.41 | 288 | 88.54% | 80.90% | 1.08E-02 |
| Sheng Ma | Fatigue | -0.40 | 65 | 92.31% | 73.85% | 5.00E-03 |
| Niu Xi | Insomnia | -0.36 | 83 | 84.34% | 71.08% | 4.02E-02 |
| Bai Xian Pi | Poor appetite | -0.33 | 32 | 81.25% | 56.25% | 3.10E-02 |
| Zhi Shi | Yellow skin | -0.21 | 124 | 54.84% | 35.48% | 2.20E-03 |
| Zhe Bei | Insomnia | -0.12 | 49 | 93.88% | 77.55% | 4.33E-02 |
| Di Ding | Fatigue | -0.02 | 78 | 91.03% | 76.92% | 1.64E-02 |
| Jing Jie | Insomnia | -0.01 | 38 | 89.47% | 60.53% | 8.07E-03 |

## VI.    Discussion

In this work we established a network medicine framework for TCM by systematically mapping symptoms-associated genes and herb targets onto the human protein interactome, and analyzing their topological relations in the network. We first observed that genes associated with a symptom are not distributed randomly on the interactome, but cluster into localized modules; furthermore, a short network distance between two symptom modules is predicative of the symptoms co-occurrence and similarity. We then show that the network proximity between a herb's targets and a symptom module is predictive of the herb's effectiveness in treating that symptom. We validate our computational analysis with real-world hospital data, showing that higher relative risk of symptoms correlates with shorter interactome distance in patients, and herb-symptom proximity predicts herb-symptom effectiveness. Finally, we identified novel herb-symptom pairs that are predicted effective by network proximity and proven effective in hospital data, but not yet recognized by the TCM community, highlighting the application value of our framework in identifying herb discovery and repurposing opportunities.

To our knowledge, this is the first attempt to study TCM on a systematic level, whereas previous research is limited to single herbs or prescriptions. We propose that the high-level TCM principles are found on symptom-herb relations from the human protein interactome perspective. We showed that analyzing the effects of TCM on symptoms is proven helpful in understanding TCM from a modern biology and modern (western) medicine perspective. Moreover, compared with existing network pharmacology approaches that rely completely on target-disease gene overlap (i.e. they use the "herb/drug directly targets disease gene" paradigm), our whole-interactome approach is more generic as we no longer assume herb/drug targets have to hit disease/symptom genes directly, but can still be effective if they hit the network neighborhood, which is the pattern we consistently observed in previous works in drug-disease relations [15, 18]. We are also the first to have designed multiple pipelines to approximate herb-chemical-target relation in our network proximity pipelines, which may lead to prioritization of effective chemicals in future work. Our data and network framework established a generic, scientific platform to systematically study TCM. Our approach combining computational network science and hospital data bioinformatics offers a powerful inter-disciplinary way to study traditional medicine and make insightful novel herb treatment predictions.

A challenge in our work is the insufficient accuracy and completeness of the available data. Herb-chemical-target data as well as symptom gene association data are always noisy and incomplete. The hospital data we used in this study is limited to Liver cirrhosis inpatients (i.e., it is not generic enough). The data is also insufficient to support large-scale bioinformatic analysis, only supporting propensity score matching on a limited subset of the data. It is also a biased dataset in the sense that all herbs that appear in this dataset are supposed to treat the disease/symptoms, thus negative samples of herbs with no effects are missing. Another challenge is the diversity/complexity in the herb-chemical-target relations, when trying to characterize herb-symptom relation. Here we used the simplest ways to define their network relations, by taking the target union of the chemicals or taking average/minimum network distance of the chemicals. More refined chemical data could improve the results; alternative metrics may better capture the network relations; or, consensus algorithms may balance the results of multiple pipelines[18]. Including more comprehensive and refined datasets and exploring more complicated methods/metrics are promising future directions.

TCM prescriptions, or herb combinations, is another interesting future direction. Our recent work and Wang et al. showed that co-prescribed herbs tend to be close in the protein interactome [47, 48]. According to the classic TCM concept, each herb in a prescription has a specific effect, which is often complementary to the effects of other co-prescribed herbs. Now that we have established the framework to study herb-symptom relations in this work, we may be able to further explore how the combination of herbs works specifically against a given symptom or symptom set in a future project.

**Methods and Data**

**Symptom-gene association data**

We used symptom data from Symmap, which integrates disease-gene association from DisGeNet[49] and MalaCards[50]. The data contain 110,407 associations with 11,362 unique diseases represented by Unified Medical language System (UMLS) concept codes and 13,271 unique genes. To obtain the high-quality symptom gene associations, we utilized the concept of "dual phenotypes" (DP)[51], such as obesity, fever, and insomnia, which are regarded as both diseases and symptoms. Thus, the symptom-gene associations are straightforwardly the corresponding disease-gene associations, for diseases with DP properties. In order to identify these kinds of phenotype terms (e.g. symptom) from databases, we filtered an integrated dual phenotype-genotype associations dataset by limiting the semantic types of UMLS concepts as symptoms from the disease-gene associations [52]. Here, we obtained 16,049 associations between 490 symptoms with concept unified identifiers (CUI) code and 4193 genes from the related previous work. To ensure the reliability of the symptom-associated genes data, we focus on the 174 symptoms with at least 20 associated genes, ignoring the symptoms with less genes as their data may be too incomplete.

**Herb, chemical and target data**

We used herb data from (1) the recently updated HIT 2.0 database [32], and (2) TCMIO database, a comprehensive collection of TCMSP, TCMID, and TCM-ID databases [33-35]. HIT database has straightforward herb target data so we use them directly. For the other TCM databases, we consider a herb as an assembly of chemicals and use their chemical composition data. The TCM databases focus on chemicals with potential therapeutic effect, rather than common chemicals. Then we obtain the protein targets of each chemical from STITCH database[53], keeping only targets with experimental evidence. In addition to herb-chemical-target data, we also used a herb-symptom indication dataset from SymMap[36], an expert-curated list of herb-symptom pairs recognized by doctors as effective treatment.

**Human Protein Interactome**

We use the human protein-protein interactome from our previous work of predicting COVID-treating drugs [18]. The interactome is assembled using experimentally validated protein interactions including: (1) binary PPIs, derived from high-throughput yeast two-hybrid experiments, three-dimensional protein structures; (2) PPIs identified by affinity purification followed by mass spectrometry; (3) kinase substrate interactions; (4) signaling interactions; and (5) regulatory interactions. The final interactome used in our study contains 18,505 proteins, and 327,924 interactions between them.

**Hospital data - Clinical symptom-herb associations**

We have collected the electronic medical record (EMR) data of Liver cirrhosis inpatient cases from Hubei Provincial Hospital of Traditional Chinese Medicine (TCM) in Wuhan, which included the full clinical profiles of patients. TCM clinical named entities, such as symptoms and their trajectory (e.g. symptom recovery) were extracted from the admission and discharge records using text-mining methods based on a clinical information extraction tool (Human-machine Cooperative Phenotypic Spectrum Annotation System, www.tcmai.org, HCPSAS). The resulting dataset contains 1936 inpatients with herb prescription records, which usually consisted of 16-18 herbs used in combination for treatment. Hence, we considered that if a prescription treats the symptoms, the herbs included in this prescription are related to the symptoms. Finally, we obtained 5106 symptom-herb associations which involve 55 symptoms and 218 herbs. All admission data of these patients were verified and standardized by the trained medical researchers to ensure highly accurate terminological mappings.

**Hospital data - Symptom terminology mapping and processing**

To connect clinical and genetic data, we manually mapped Chinese terms of symptoms and herbs in clinical data to English terms in symptom-gene associations by trained medical researchers, thereby ensuring highly accurate terminological mappings. 315 English symptom terms with associated genes mapped to 92 Chinese symptom terms in clinical data. Therefore, there is a phenomenon of multiple UMLS code merging corresponding to one TCM symptom, for example, C0277799 and C0015967 were both mapped to 发热 (fever).

**Hospital data - Propensity Score Matching**

We used Propensity Score Matching (PSM) in the clinical dataset to remove the biases of patient basic information on herb treatment outcomes. PSM is a statistical matching technique that attempts to estimate the effect of a treatment, policy, or other intervention by accounting for the covariates that predict receiving the treatment[54]. For a designated herb-symptom pair, we matched the patients with the herb-symptom pair against other herbs treating the same symptom, to evaluate the effectiveness of the herb-symptom pair. For example, for the *Baizhu*-fatigue pair, the fatigue patients who received *Baizhu* therapy at any point duration hospitalization were defined as the case group of *Baizhu*-fatigue pair. Fatigue patients that did not receive *Baizhu* treatment form the control group. We adjusted for baseline characteristics (e.g. age, sex) and high-incidence comorbidity characteristics of patients in the two groups. The most common comorbidities selected to control include esophageal and gastric varices, abdominal effusion, hypoproteinemia, hypertension, and diabetes. In the propensity score matching analysis, the nearest-neighbor method was applied to create a 1:1 matched control sample.

## Metrics

### LCC and LCC z-score

We characterize the localization of a node set in the network with the z-score of the large-connected-component (LCC) [14]. We first compute the size of the LCC formed by the node set, and then compare the observed LCC size against the random expectation generated from simulations preserving degree of the nodes[15]. The LCC z-score is the difference between the observed LCC size and the mean of randomization $\mu(random\ Lcc)$, divided by standard deviation of the randomization $\sigma(random\ Lcc)$:

$$z_{LCC} = \frac{Observed\ Lcc\ size - \mu(random\ Lcc)}{\sigma(random\ Lcc)},$$

An LCC z-score larger than 1.6 indicates the observed LCC is significantly larger than random expectation.

An implementation of the code for LCC and its z-score computation can be found in [14].

### Network separation

We measure the network relation between two node sets (e.g. target modules of herbs A and B) using the network separation metric that successfully characterized disease-disease relation and drug-drug relation previously [14, 16]:

$$s_{AB} = \langle d_{AB} \rangle - \frac{\langle d_{AA} \rangle + \langle d_{BB} \rangle}{2}$$

The network separation metric compares the mean shortest distance within the interactome between the nodes of each node set, $\langle d_{AA} \rangle$ and $\langle d_{BB} \rangle$, to the mean shortest distance $\langle d_{AB} \rangle$ between node sets A and B. In $\langle d_{AB} \rangle$, targets associated with both herb A and B have a zero distance by definition. The random expectation of $s_{AB}$ is zero. A negative $s_{AB}$ means the two node sets are located in the same network neighborhood; while a positive $s_{AB}$ means the two node sets are topologically separated.

An implementation of network separation computation can be found in [14].

### Symptom semantic similarity

In order to evaluate the biological similarity between a pair of symptoms, we use semantic similarity [25] to characterize the biological similarity of genes associated with the symptoms.

We used the python package pygosemsim ([https://github.com/mojaie/pygosemsim](https://github.com/mojaie/pygosemsim)) to compute the GO (Gene Ontology) semantic similarity between a pair of genes. For semantic similarity of two symptoms, we compute the average GO semantic similarity of all pairs of genes between the symptoms.

**Network proximity distance and z-score**

Given $S$, the set of symptom-associated genes, $T$, the set of herb targets, and $d$ (s, t), the shortest path length between nodes $s \in S$ and $t \in T$ in the network, we define the network proximity distance metric (referred to as "proximity distance d" in the main text) as the average distance over targets to their closest symptom-associated gene[15]:

$$d(V, T) = \frac{1}{||T||} \sum_{t \in T} \min \ d(v, t)$$

Then we convert this absolute distance $d$ to a relative proximity z-score, by simulating the random expectation of distances between two randomly selected groups of proteins, matching the size and degrees of the original $V$ and $T$ sets. To avoid repeatedly selecting the same high degree nodes, we use degree-binning[15]. Using the mean $\mu(V, T)$ and standard deviation $\sigma(V, T)$ of the simulated reference distribution, we define the network proximity z-score as:

$$z(V, T) = \frac{d(V, T) - \mu(V, T)}{\sigma(V, T)}$$

The proximity z-score measures how the proximity distance differs from random expectation, with z=0 being neutral, z<0 being more proximal than random, and z>0 being more distant from random. For both proximity distance d or proximity z-score, the lower the metric value, the closer the two node sets are on the network. Note the proximity z-score is a stochastic measure because of the randomized simulation, i.e. identical repeated computation don't yield identical z-scores. An implementation of network proximity metrics computation can be found in [15] or [18].

**Herb-target mapping methods to obtain herb-symptom distance for each herb-symptom pair**

Four herb-target mapping methods were deployed to obtain herb-symptom distance $d_{hs}$, from (a) HIT direct target data, and (b-d) herb-chemical-target data.

Notations: $h$: herb, $s$: symptom-associated genes, $c_i$: the $i_{th}$ composing chemical of a herb, $T$: targets of a herb or chemical, $proximity(T, s)$: proximity measures calculated from $(T, s)$, i.e. proximity $d(V, T)$ or $z(V, T)$.

In the cases where herb targets are directly associated, i.e. (a) the HIT database or (b) Target Union, the herb-symptom distance metric is, straightforwardly, the proximity metric(s) between herb targets and symptom genes as $d_{hs} = proximity(T_h, s)$. In (a) HIT data $T_h$ is given directly; in (b) the targets of a herb is defined as the union of targets from all the herb's composing chemicals: $T_h = \bigcup_{c_i \in h}(T_{c_i})$.

When herb targets are not directly available in (c, d), we define 2nd order herb-symptom distance metrics from 1st order chemical-symptom distances. The 1st order chemical-symptom distance is the proximity distance or z-score for a chemical and a symptom, using targets of this chemical, denote as $d_{c_i s} = proximity(T_{C_i}, s)$. Then based on this 1st order distance, we define the 2nd order herb-symptom distance b as the (c) average or (d) minimum of all 1st order distances:

(c) Average: $d_{hs}^{avg} = \frac{\sum_{c_i} d_{c_i s}}{N_c}$, $N_c$ being the total number of chemicals in this herb

(d) Minimum: $d_{hs}^{min} = \min_{c_i \in h} d_{c_i s}$

Together, these four herb-target mapping methods crossing the two proximity measures yields 8 herb-symptom proximity pipelines.

**Supplementary data files**

- **We will provide these data files <u>after</u> the final publication of the manuscript.**

S1. Symptom gene association, symptom by UMLS ID, filtered for symptoms with >= 20 associated genes

S2. Human protein Interactome

S3. Symptom pairwise network distance $D_{ab}$ & $S_{ab}$ along with co-disease count, and GO semantic similarity

S4. Herb-chemical-target data:

  1. HIT target data, mapped to herb UID and protein entrez ID, and filtered for targets within the human interactome

  2. TCM herb-chemical composition data directly obtained from TCMIO

  3. Chemical-target data, processed from STITCH, removing common chemicals from Foodome data and filtered for targets within the human interactome

S5. Network proximity metrics from 8 pipelines across the TCM databases, with indication annotation:

  1. P1&P2: HIT database

  2. P3&P4: Target Union

  3. P5-P8: 2nd order distances

S6. Symptom associated genes and symptom co-occurrence data

S7. Herb-symptom proximity, for all 8 pipelines, with an Indicator whether it is in hospital data, and an Indicator for effectiveness

S8. PSM effectiveness result of 886 herb-symptom pairs, with 8 network metrics

S9. Herb name-ID mapping

**Acknowledgement**

This work is supported by fundings of the Barabási Lab, the Zhou Lab, and the Albert Lab.

The authors thank Dr. Michael Sebek, Jiali Cheng, Dr. Italo do Valle, Dr. Deisy M. Gysi, Rui Hua for help discussions.